\documentclass[aps,12pt,final,notitlepage,oneside,onecolumn,nobibnotes,nofootinbib,
superscriptaddress,noshowpacs,centertags]{revtex4}
\usepackage{graphicx}
\begin{document}

\title{PSEUDOSPIN SYMMETRY AND STRUCTURE OF NUCLEI WITH $Z\ge$ 100}

\author{\firstname{R. V.}~\surname{Jolos}}
 \email{jolos@theor.jinr.ru}
 \affiliation{Bogoliubov Laboratory of Theoretical Physics, JINR, 141980 Dubna, Russia}
\author{\firstname{V. V.}~\surname{Voronov}}
 \email{Voronov@theor.jinr.ru}
 \affiliation{Bogoliubov Laboratory of Theoretical Physics, JINR, 141980 Dubna, Russia}

\begin{abstract}
\vspace*{.5cm} In the framework of the Relativistic Mean Field
Approach a pseudospin dependence of the residual forces in nuclei
is considered. It is shown that this dependence is relatively
weak. As a consequence, a pseudospin dependence of the
particle--core coupling is weak as well. This leads to a small
splitting of the pseudospin doublets produced by a vector coupling
of an odd particle pseudospin and a pseudo--orbital momentum of
the core. Some possibilities for experimental investigations of
the manifestations of the pseudospin symmetry in the spectra of
odd nuclei with $Z\ge$ 100 are indicated.
\end{abstract}

\maketitle

\section{Introduction}

The pseudospin symmetry \cite{Arima,Hecht,Ginocchio} is known as
approximate symmetry of the nuclear mean field. This symmetry is
manifested in the nuclear excitation spectra by the presence of
quasi--degenerate doublets. At the same time, the  existence of
this symmetry is strongly related to the strength of the
spin--orbit interaction term of the nuclear mean field, and
therefore, to the next proton magic number. The strong spin--orbit
interaction in nuclei and the presence of  approximate pseudospin
symmetry in a nuclear mean field are two sides of the same medal.

It is well known that the mean field in nuclear theory plays a
role of a basic theory for several more specific advanced
theories. These theories can be built upon  introducing the single
particle mean field basis. Therefore, it is very important for the
whole field of nuclear structure physics to examine consequences
of the pseudospin symmetry, although this symmetry is approximate.

Any dynamical symmetry implies the existence of a characteristic
multiplet structure. These multiplets are characterized by a
magnitude of the multiplet splitting. The characteristic magnitude
of a splitting of the pseudospin doublets in spherical nuclei is
of the order of 0.1$\hbar\omega_0$  where $\hbar\omega_0$ is a
frequency of the single particle oscillator.

However, this splitting demonstrates a dependence on a ratio
between the numbers of protons and neutrons and it is very small
in some nuclei.

Single particle pseudospin doublets in deformed nuclei are
characterized by a projection of the pseudo--orbital momentum on
the symmetry axis. The splittings of these doublets are several
times smaller than in spherical nuclei. The doublet structure is
also observed in the rotational bands of odd deformed nuclei based
on the pseudospin singlets, i.e., on the states with the
projection of the pseudo--orbital momentum on the symmetry axis
equal to zero. The doublet structure in these bands arises as a
result of coupling of an odd particle pseudospin and a total
pseudo--orbital momentum. This pseudo--orbital momentum is a sum
of a core rotational momentum and a pseudo--orbital momentum of an
odd particle. A splitting of these doublets is quite small and
equals  several tens of KeV.

These facts mean that the term describing a particle--core
coupling in a phenomenological nuclear Hamiltonian is pseudospin
independent with good accuracy. For this reason, the spectra of
odd deformed nuclei and especially rotational bands based on the
pseudospin singlets are the most interesting objects to look for
pseudospin symmetry manifestations.

The calculations performed in \cite{RingArima} have shown that the
goodness of the pseudospin symmetry improves when a nucleon
binding energy decreases and a pseudo--orbital momentum decreases.
Therefore, weakly bound exotic nuclei are the most exciting ones
to search for the pseudospin symmetry manifestation.

It is the aim of the present paper to investigate a pseudospin
dependence of the particle--core coupling and indicate  some
experimental possibilities to study  the pseudospin symmetry
manifestation in the spectra of  odd superheavy nuclei.

\section{Pseudospin symmetry origin}

A strong spin--orbit interaction introduced into nuclear physics
in 1949 \cite{shellmodel} was an unusual idea at that moment as
the majority of nuclear physicists believed in the $L-S$ coupling
scheme. However, strong spin--orbit interaction was necessary to
reproduce the known magic numbers. The simplest Hamiltonian which
can  describe the nuclear mean field is the Hamiltonian with a
harmonic oscillator potential, spin--orbit and orbit--orbit terms
\begin{eqnarray}
\label{osc}
h=h_{osc} + \nu_{ls}{\bf l}\cdot{\bf s} + \nu_{ll}\left({\bf l}^2 - \langle{\bf l}^2\rangle_{shell}\right)
\end{eqnarray}
The value of $\nu_{ls}$ is such that a splitting generated by the
${\bf l}\cdot{\bf s}$ term in (\ref{osc}) is large.

Twenty years later \cite{Arima,Hecht} quasidegeneracy in the
single particle level scheme was observed. Namely, single particle
states with $j_1 = l_1 + 1/2$ and $j_2 = l_2 (=l_1 + 2) - 1/2 =j_1
+ 1$ lie very close in energy. They are labeled as pseudospin
doublets with the following quantum numbers:
\begin{eqnarray}
\label{pseudo1}
\tilde{N}&=& N-1\nonumber\\
\tilde{l}&=&\left\{\begin{array}{rcl}l_1 + 1, j_1=l_1 + 1/2\\
l_2 - 1, j_2 = l_2 - 1/2\\\end{array}\right.\nonumber\\
\tilde{s}&=&1/2,
\end{eqnarray}
where tilde marks the pseudo--oscillator quantum numbers. Examples of pseudospin doublets are:
$3s_{1/2}$ and $2d_{3/2}$ ($\tilde{l}=1, \tilde{N}=3$), $1g_{7/2}$ and $2d_{5/2} (\tilde{l}=3, \tilde{N}=3$),
$1h_{9/2}$ and $2f_{7/2} (\tilde{l}=4, \tilde{N}=4$). An example of a pseudospin singlet is $3p_{1/2} (\tilde{l}=0, \tilde{N}=4$).

In terms of the pseudospin--orbit operators the Hamiltonian (\ref{osc}) takes the form \cite{Bahri}

\begin{eqnarray}
\label{pseudoosc} h=\tilde{h}_{osc} + (4\nu_{ll}-\nu_{ls}){\bf
{\tilde{l}}}\cdot{\bf{\tilde{s}}} +
\nu_{ll}\left({{\bf{\tilde{l}}}}^2-\langle{\bf{\tilde{l}}}^2\rangle_{shell}\right)
+ const.
\end{eqnarray}

It is known empirically that
\begin{eqnarray}
\label{pseudoosc1}
4\nu_{ll}-\nu_{ls}\approx 0.
\end{eqnarray}
As a result, pseudospin--orbit interaction is several times weaker
than usual spin--orbit interaction.

The physical grounds for appearance of the pseudospin symmetry in
nuclei was clarified in the works of J.~N.~Ginocchio
\cite{Ginocchio}. It was shown that this problem should be
considered in the framework of the Relativistic Mean Field Theory.
The Lorentz covariant Dirac equation for a single particle with
mass $M$ is
\begin{eqnarray}
\label{dirac1}
\left(\gamma^{\mu}(cp_{\mu}+g_v A_{\mu})+Mc^2+V_s\right)\Psi =0,
\end{eqnarray}
where $V_s$ is a scalar potential, attractive in the case of nucleons, and $A_{\mu}(A_0 ,{\bf A})$ is a vector potential.
Assuming that these potentials are time independent we obtain  a Dirac Hamiltonian
\begin{eqnarray}
\label{dirac2}
H={\bf \alpha}\left(c{\bf p} + g_v{\bf A}\right) + V_v + \beta\left(Mc^2+V_s\right),
\end{eqnarray}
where $V_v =g_v A_0$ is repulsive \cite{Ring}. Neglecting ${\bf
A}$ which is not presented in a mean field \cite{Ring} of an
even--even nucleus we obtain the following equation for the large
($g$) and small ($f$) components of the Dirac spinor \cite{Ring}
\begin{eqnarray}
\label{dirac3}
\left(\begin{array}{cc}
M+V_v+V_s & {\bf  \sigma}\cdot{\bf p}\\
{\bf  \sigma}\cdot{\bf p} & -M+V_v-V_s\end{array}\right)\left(\begin{array}{c}g\\f\end{array}\right)=E\left(\begin{array}{c}g\\f\end{array}\right)
\end{eqnarray}
Representing $E$ as $E=M+\epsilon$ and using the fact that
$|\epsilon |\ll 2\tilde{M}$ where $\tilde{M}=M-1/2(V_v - V_s)/c^2$
we derive from (\ref{dirac3}) the Schr\"odinger equation for the
large component $g$ \cite{Ring}
\begin{eqnarray}
\label{dirac4}
\left({\bf p}\frac{1}{2\tilde{M}}{\bf p} + \frac{\hbar^2}{4\tilde{M}^2c^2}\frac{1}{r}\frac{\partial (V_v - V_s)}{\partial r}{\bf l}\cdot {\bf s}+(V_v + V_s)\right)g=\epsilon g.
\end{eqnarray}
It is seen from (\ref{dirac4}) that different combinations of
$V_v$ and $V_s$ contribute to the spin--orbit term and the radial
potential well. The depth of the radial potential $(V_v + V_s)$ is
equal approximately to 50 MeV. The value of $(V_v - V_s)$ is equal
to 700--800 MeV inside the nucleus.

As was shown in \cite{Ginocchio}, pseudospin symmetry takes place
if $V_s$/$V_v=-1$. For this reason, this symmetry is not exact
because in this case $(V_v + V_s)$=0, i.e., there is no binding
potential for nucleons. However, as it follows from the QCD sum
rule, the ratio $V_s$/$V_v\approx -1$ with an estimated accuracy
of 20$\%$. Indeed the detailed QCD sum rule gives
\begin{eqnarray}
\label{QCD1}
V_s &=& -4\pi^2\sigma_N\rho_N/M^2 m_q,\nonumber\\
V_v &=& 32\pi^2\rho_N /M^2,
\end{eqnarray}
where $\rho_N$  is the nuclear matter density, $\sigma_N$ is the
so-called sigma term ($\sigma_N\approx 45\pm 8$ MeV) and $m_q$ is
the mass of a light quark. Thus,
\begin{eqnarray}
\label{QCD2}
\frac{V_s}{V_v}=-\frac{\sigma_N}{8m_q}\approx -1.1.
\end{eqnarray}

\section{Pseudospin dependence of the particle--core coupling}

To describe the properties of the low--lying collective states and
of a coupling of a single particle and a collective motion, it is
useful to replace a realistic residual interaction by a schematic
interaction. A useful application of this concept is the RPA
\cite{Rowe}. The RPA is equivalent to the Time Dependent
Hartree--Fock. It means that in the framework of the RPA all
interactions generating the same time--dependent mean field are
equivalent. For the time--dependent mean field $U({\bf r},t)$ we
have a relation
\begin{eqnarray}
\label{7.0}
U({\bf r},t)=\int d^3r' V_{res}({\bf r},{\bf r}')\rho ({\bf r}',t),
\end{eqnarray}
where $\rho$ is a nuclear density and $V_{res}$ is a residual
interaction.  Having in mind a description of nuclear shape
oscillations and their coupling to a single particle motion let us
parameterise a time dependence of the mean field and the nuclear
density by the following expressions:
\begin{eqnarray}
\label{7.1}
U({\bf r},t)=U_0\left(\frac{r}{1+\sum_{\lambda ,\mu}\alpha_{\lambda ,\mu}(t)Y_{\lambda ,\mu}({\bf r})}\right),
\end{eqnarray}
\begin{eqnarray}
\label{7.2}
\rho({\bf r},t)=\rho_0\left(\frac{r}{1+\sum_{\lambda ,\mu}\alpha_{\lambda ,\mu}(t)Y_{\lambda ,\mu}({\bf r})}\right),
\end{eqnarray}
where $\rho_0$ and $U_0$ are the static mean field and density.
Expanding in powers of $\alpha_{\lambda ,\mu}$ in Eqs. (\ref{7.1})
and (\ref{7.2}), restricting ourselves to the first orders in
$\alpha_{\lambda ,\mu}$
\begin{eqnarray}
\label{8.1}
U({\bf r},t)=U_0(r)-r\frac{d\rho_0(r)}{dr}\sum_{\lambda ,\mu}\alpha_{\lambda ,\mu}(t)Y_{\lambda ,\mu},
\end{eqnarray}
\begin{eqnarray}
\label{8.2}
\rho({\bf r},t)=\rho_0-r\frac{dU_0(r)}{dr}\sum_{\lambda ,\mu}\alpha_{\lambda ,\mu}(t)Y_{\lambda ,\mu}
\end{eqnarray}
and substituting the result into (\ref{7.0}) we obtain $V_{res}$ in a separable form
\begin{eqnarray}
\label{8.3}
V_{res}({\bf r},{\bf r}')=\chi r\frac{dU_0(r)}{dr}\cdot r'\frac{dU_0(r')}{dr'}\sum_{\lambda ,\mu }Y_{\lambda ,\mu }({\bf r})Y_{\lambda ,\mu }^*({\bf r}')
\end{eqnarray}
with the condition for $\chi$
\begin{eqnarray}
\label{8.4}
1=\chi\int r^2 dr r\frac{dU_0}{dr}\cdot r\frac{d\rho_0}{dr}.
\end{eqnarray}

For the nuclear mean field $U_0$ the following expression results
from  Eq.(\ref{dirac4}):
\begin{eqnarray}
\label{mf1}
U_0(r)=\left(V_s(r)+V_v(r)\right)+\frac{\hbar^2}{4\tilde{M}^2c^2}\frac{1}{r}\frac{d}{dr}\left(V_s(r)-V_v(r)\right)({\bf l}\cdot{\bf s}).
\end{eqnarray}
Let us approximate $V_s(r)$ and $V_v(r)$ by the terms linear in $\rho (r)$ \cite{Weise}
\begin{eqnarray}
\label{mf2}
V_s(r)&=&-V_{0s}\frac{\rho_0(r)}{\rho_{av}},\nonumber\\
V_v(r)&=&V_{0v}\frac{\rho_0(r)}{\rho_{av}},\nonumber\\
V_{0s}-V_{0v}&\approx& 50MeV,\nonumber\\
V_{0s}+V_{0v}&=&700\div 800 MeV,\nonumber\\
\tilde{M}&=&M-(V_v-V_s)/2c^2,
\end{eqnarray}
where $\rho_{av}$ is the nuclear density inside the nucleus. Then,
assuming a Saxon--Woods form of $\rho_0$ we obtain the following
expression for the formfactor $r\frac{dU_0(r)}{dr}$:
\begin{eqnarray}
\label{9.1}
r\frac{dU_0(r)}{dr}=\frac{\rho_0(r)}{\rho_{av}}\left(1-\frac{\rho_0(r)}{\rho_{av}}\right)\left\{(V_{0s}-V_{0v})\frac{r}{a}
+ \frac{\hbar^2}{2Ma^2}\frac{{\bf l}\cdot{\bf s}}{\left(1-\frac{V_{0v}+V_{0s}}{2Mc^2}\frac{\rho_0(r)}{\rho_{av}}\right)^3}\right.\nonumber\\
\times\left [\frac{(V_{0v}+V_{0s})}{2Mc^2}\left(1-\frac{V_{0v}+V_{0s}}{2Mc^2}\frac{\rho_0(r)}{\rho_{av}}\right)\frac{a}{r}+\frac{(V_{0v}+V_{0s})}{2Mc^2}(1-2\frac{\rho_0(r)}{\rho_{av}})\right.\nonumber\\
\left.\left. +\left(\frac{V_{0v}+V_{0s}}{2Mc^2}\right)^2\frac{\rho_0(r)}{\rho_{av}}\right]\right\},
\end{eqnarray}
where $a$ is a diffusion parameter of the nuclear density
$\rho_0$. The function
$\frac{\rho_0(r)}{\rho_{av}}\left(1-\frac{\rho_0(r)}{\rho_{av}}\right)$
is localized at the nuclear surface. So we can put in the figure
brackets in (\ref{9.1}) $r=R$ where $R$ is a nuclear radius.
Therefore, we can approximate the formfactor $r\frac{dU_0}{dr}$ by
the expression
\begin{eqnarray}
\label{9.1a}
r\frac{dU_0}{dr}=\frac{\rho_0(r)}{\rho_{av}}\left(1-\frac{\rho_0(r)}{\rho_{av}}\right)\left(c+b+b({\bf l}\cdot{\bf s})\right)
\end{eqnarray}
or in terms of the pseudospin and pseudo--orbital momentum operators as
\begin{eqnarray}
\label{9.2}
r\frac{dU_0}{dr}=\frac{\rho_0(r)}{\rho_{av}}\left(1-\frac{\rho_0(r)}{\rho_{av}}\right)\left(c-b(
\bf {\tilde{l}}\cdot\bf {\tilde{s}})\right)
\end{eqnarray}
Using the values of $V_s$ and $V_v$ given above and putting $R$=7
fm, $a$=0.6 fm we obtain $c\approx 550$ MeV and $b\approx 45$ MeV.
With the formfactor (\ref{9.2}) substituted into (\ref{8.3}) the
residual forces obtained looks like the Surface Delta Interaction
\cite{Moszkowski}.

From a comparison of the values of the parameters $c$ and $b$ we
can see that the main part of the residual interaction is
pseudospin independent. Therefore, the Hamiltonian with the
pseudospin symmetric mean field term and the pseudospin
independent part of the residual forces derived above can be
considered as an approximate model for description of low--lying
nuclear excitations. In the framework of this model the excited
states of both even--even and odd nuclei will be characterized by
the total pseudo--orbital momentum. The pseudo--orbital momentum
is finally coupled to the pseudospin forming pseudospin
multiplets. The eigenstates of this Hamiltonian with one sort of
particles belong to the basis of irreducible representations of
$U(\Omega )\bigotimes U(2)$. Here $\Omega$ is the total number of
the pseudo--orbital $\tilde{m}$--states. The spatial parts of the
nucleon wave functions form the basis of irreducible
representations of $U(\Omega )$ characterized by their symmetry
type. A more detailed characterization of these states could be
provided by a subgroup of $U(\Omega )$ containing $O(3)$. In the
case of  well-deformed nuclei with their rotational bands as basic
elements of the excitation spectra the intermediate group is
$SU(3)$. In the case of even--even nuclei the lowest bands
correspond to the most symmetric representation characterized by
the following sequence of  values of the pseudo--orbital momenta:
$\tilde{L}$=0,2,4,... In the case of odd nuclei, the value of the
pseudo--orbital momentum of the lowest state can be different from
zero. This value depends on the pseudo--orbital momenta of the
single particle state near the Fermi surface.

A splitting of the pseudospin multiplets is determined by the matrix element of the pseudospin dependent
part of the residual forces $\delta V_{res}$. In the first approximation
\begin{eqnarray}
\label{11.1}
\delta V_{res}({\bf r_1},{\bf r_2})=-2c b\chi\left({\bf{\tilde{l}_1}}\cdot{\bf{\tilde{s}_1}}\right)\frac{\rho_0(r_1)}{\rho_{av}}\left(1-\frac{\rho_0(r_1)}{\rho_{av}}\right)\nonumber\\
\times \frac{\rho_0(r_2)}{\rho_{av}}\left(1-\frac{\rho_0(r_2)}{\rho_{av}}\right)\sum_{\lambda ,\mu}Y_{\lambda ,\mu}({\bf r}_1)Y^*_{\lambda ,\mu}({\bf r}_2).
\end{eqnarray}
In the lowest states pseudospin takes the minimum possible value
since in this case the coordinate depending part of the wave
function is the most symmetric and the pseudospin of the nucleon
pairs is zero. Thus, the total pseudospin is equal to that of an
odd particle, i.e., to 1/2. In this case  index "1" in
(\ref{11.1}) belongs to an odd particle, but index "2" describes
all other particles forming the core. Then the interaction term
(\ref{11.1}) takes the form of the particle -- core coupling term
considered, for example, in the Bohr--Mottelson model. The
pseudospin independent analog of the Bohr--Mottelson
particle--core coupling term with the radial formfactor derived
above is
\begin{eqnarray}
\label{11.1a}
\delta V_{res}({\bf r_1},{\bf r_2})=-c^2\chi\frac{\rho_0(r_1)}{\rho_{av}}\left(1-\frac{\rho_0(r_1)}{\rho_{av}}\right)\nonumber\\
\times \frac{\rho_0(r_2)}{\rho_{av}}\left(1-\frac{\rho_0(r_2)}{\rho_{av}}\right)\sum_{\lambda ,\mu}Y_{\lambda ,\mu}({\bf r}_1)Y^*_{\lambda ,\mu}({\bf r}_2).
\end{eqnarray}
Comparing (\ref{11.1}) and (\ref{11.1a}) we can see that the
strength of the interaction term (\ref{11.1}) is
2$\frac{b}{c}\langle {\bf{\tilde l}}_i\cdot{\bf{\tilde
s}}_i\rangle$ times smaller than that of the pseudospin
independent particle--core coupling term. Let us estimate at first
the average $\langle {\bf{\tilde l}}_i\cdot{\bf{\tilde
s}}_i\rangle$ where index "i" denotes an odd particle. The
pseudo--orbital momentum of an odd particle contributes to the
total pseudo--orbital momentum $\tilde{L}$ of the state. It can be
taken to be equal to $\frac{1}{N}\tilde{L}$ where $N$ is
approximately the number of particles in the open shell if the
number of nucleons contributing to the total pseudo--orbital
momentum does not depend on $\tilde{L}$. If this number increases
proportionally to $\tilde{L}$ the contribution of the odd particle
to $\tilde{L}$ can be estimated as $\frac{\tilde{l}_0}{N}$ where
$\tilde{l}_0$ is a constant of the order of unity. For
well-deformed nuclei we can take $N\approx$30. Since
$\tilde{s}=$1/2, we obtain
\begin{eqnarray}
\label{12.1}
2\frac{b}{a}\langle {\bf{\tilde l}}_i\cdot{\bf{\tilde s}}_i\rangle\approx\frac{\tilde{L}}{360}\quad\mbox{or}\quad\frac{\tilde{l}_0}{360}.
\end{eqnarray}
The matrix element of the particle--core interaction term in the
Bohr--Mottelson model in the case of deformed nuclei can be
estimated as $\sim$2.3 MeV. Thus, a splitting of the pseudospin
doublets in the rotational bands of odd nuclei is
$\sim$7$\tilde{L}$ or $\sim$7$\tilde{l}_0$ KeV. This estimate is
in a correspondence with an experimentally observed splitting
which is equal to $10\div 30$ KeV; though, without proportionality
to $\tilde{L}$. Thus, we should use an estimate with
$\tilde{l}_0$.

\section{What is interesting to observe}

As it was mentioned above, experimental data on the single
particle spectra of the stable well-investigated nuclei show that
pseudospin symmetry is fulfilled only approximately. It is known
from the consideration in the framework of the Relativistic Mean
Field Approach that pseudospin symmetry improves as the binding
energy of nucleons decreases. The calculations \cite{RingArima}
have also shown that pseudospin symmetry improves as the
pseudoorbital momentum $\tilde{l}$ decreases. Therefore, it is
interesting to look for the manifestations of pseudospin symmetry
in nuclei far removed from the valley of stability. Thus, the
low--lying states in nuclei with $Z\ge 100$ are interesting
objects for investigations. In these nuclei  more interesting are
single particle states with a small pseudo--orbital momentum, or
in the case of deformed nuclei with a small projection of the
pseudo--orbital momentum on the axial symmetry axis.

At the same time, calculations of the single particle spectra of
superheavy nuclei performed up to now demonstrate different
results: from small to large splitting of the pseudospin doublets.
This is an additional argument to carry out experimental
investigations in order to clarify the problem.

The experimental data on well-investigated nuclei \cite{Jolos} and
the consideration in the previous section have shown that the most
clear manifestation of the pseudospin symmetry is expected in the
spectra of the low--lying rotational bands of odd nuclei based on
the pseudospin singlets or on the pseudospin doublets with a
projection of the pseudo--orbital momentum on the axial symmetry
axix $\Lambda$ equal to 1. Below we consider this suggestion in
detail.

The calculations of A.Parchomenko and A. Sobiczewski
\cite{Sobiczewski} show that with a large probability in odd Md
and Lr isotopes the ground state or one of the low--lying states
is [521]1/2$^-$. This state is the pseudospin singlet state having
the following pseudo--oscillator quantum numbers
$\widetilde{[420]}$1/2$^-$. The rotational band based on this
state consists of a singlet and a sequence of doublets: 1/2$^-$;
(3/2$^-$,5/2$^-$); (7/2$^-$,9/2$^-$)... For illustration of
possible observatioins let us consider nuclei with odd numbers of
neutrons $N$ equal to 101 and 103 and even numbers of protons.
Their excitation spectra are shown in Figs. 1 and 2 for
$^{173}$Hf$_{101}$, $^{171}$Yb$_{101}$ and
$^{179-185}$Pt$_{101-107}$. A small splitting of doublets equal to
several tens of KeV is the main signature of the pseudospin
symmetry. As it is seen from Figs. 1 and 2, the doublet structure can
be seen in several isotopes of the same element and for several
values of $Z$. Thus, it is not necessary to search for one very
special nucleus in which this effect is pronounced. If this effect
exists, it should be seen in several neighbouring nuclei.

The other interesting possibility for observation of the
pseudospin symmetry effects is related to the spectra of odd
isotopes of the element $Z$=111. As it is shown by the
calculations of A.Parchomenko and A. Sobiczewski, the pseudospin
doublet with $\Lambda$=1: [512]3/2$^-$ and [510]1/2$^-$ can exists
in these nuclei. The expected spectra of the low--lying states in
this case can be similar to that observed in $^{187}$Os$_{111}$
having the number of neutrons equal to 111. This spectrum is shown
in Fig. 3. It is seen that a splitting of the states in doublets
is very small and does not exceed 10 KeV.

The last example considered is the low--lying spectra of the
spherical nuclei with $Z$=115 and 117. Single particle states with
pseudo--orbital momenta $\tilde{l}$=0 and 2 can be located in
these nuclei near the ground state. The low--lying spectra should
be similar to the spectrum of $^{195}$Pt$_{117}$ which was very
well investigated in \cite{Metz}.

\section{Conclusion}

The existence of the approximate pseudospin symmetry is supported
by the experimental data for nuclei belonging to the traditionally
investigated region of the nuclide chart.

The pseudospin symmetry is justified theoretically and has its
grounds in an approximate equality of the scalar and vector
potentials in the Dirac equation describing a motion of nucleons
in a relativistic mean field.

It is interesting, whether the pseudospin symmetry will be
confirmed by  experimental data for exotic nuclei, for instance,
for superheavy nuclei.

\section*{Acknowledgements}

This work was supported in part by the RFFI grant 04--02--17376 and by the Heisenberg-Landau Program.

\newpage

\begin{figure*}[t!]
\caption{Ground state rotational bands of $^{173}$Hf and
$^{171}$Yb based on the pseudospin singlet states. Experimental
data are taken from \cite{Firestone}}
\end{figure*}

\begin{figure*}[t!]

\caption{Ground state rotational bands of $^{179,181,183}$Pt based on the pseudospin
singlet states. Experimental data are taken from \cite{Firestone}}
\end{figure*}

\begin{figure*}[t!]

\caption{The lowest--lying  rotational band of $^{187}$Os  based on the
single particle states belonging to the pseudosp[in doublet with $\tilde{\Lambda}$=1.
Experimental data are taken from \cite{Firestone}}
\end{figure*}

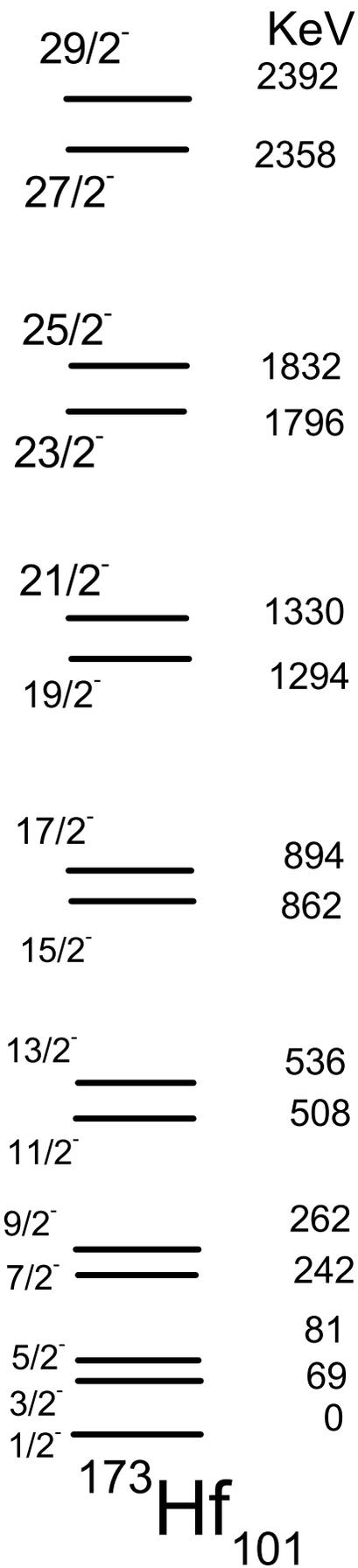
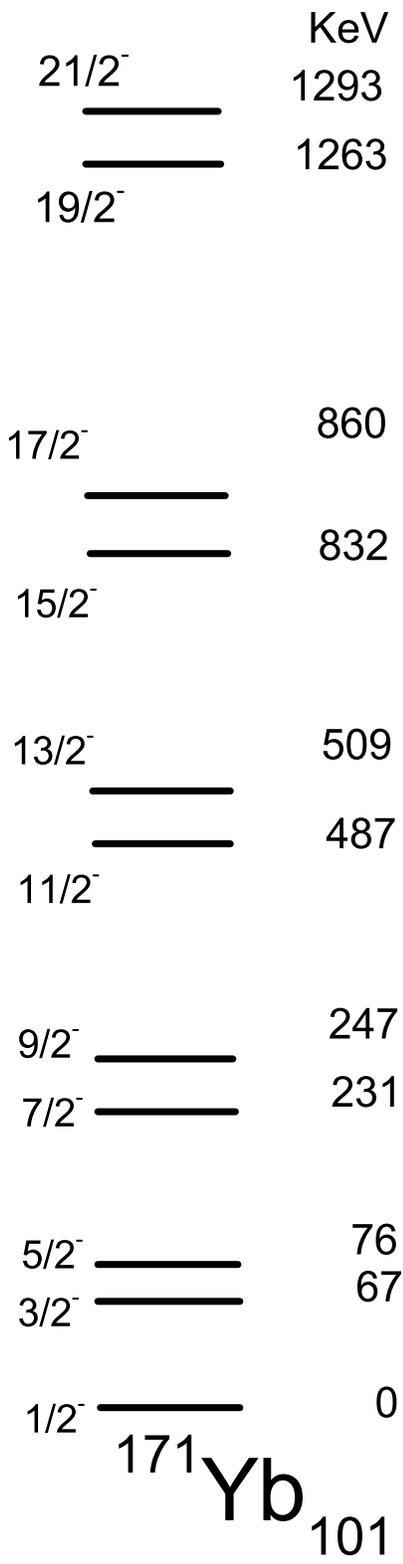
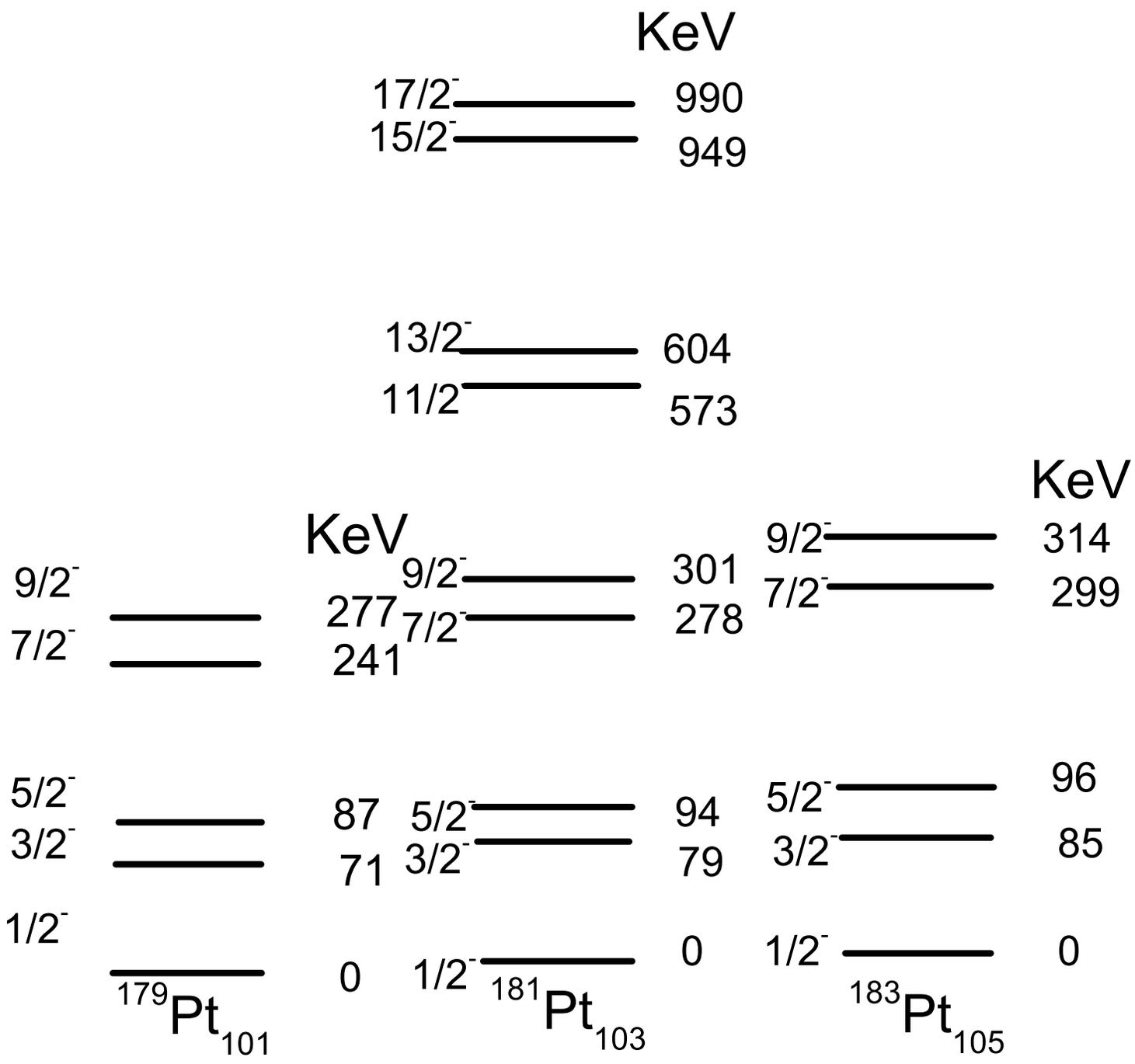
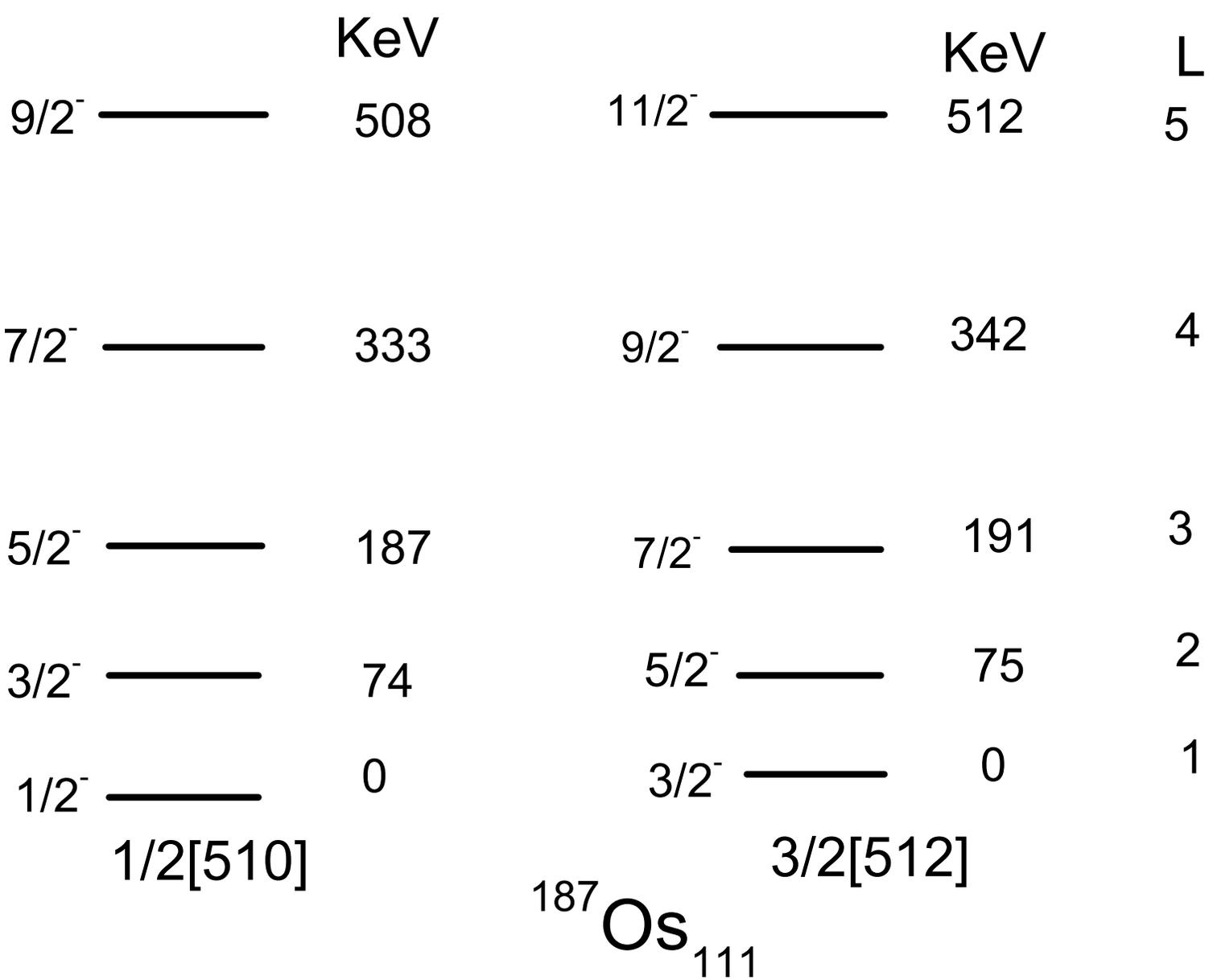
\end{document}